\documentstyle[floats,preprint,aps]{revtex}
\input epsf.sty
\newcommand{\bkg}{\mbox{$B\rightarrow K^* \gamma$}}
\newcommand{\dkev}{\mbox{$D \rightarrow K^* \ell\nu$}}
\newcommand{\vts}{\mbox{$|V_{ts}|$}}

\newcommand{\etal}{{et~al.}}
\newcommand{\vp}{{v^\prime}}
\newcommand{\pp}{{p^\prime}}
\newcommand{\vslash}{{v\!\!\!/}}
\newcommand{\Pslash}{{P\!\!\!/}}
\newcommand{\esslash}{{\epsilon\!\!/}^*}
\newcommand{\epslash}{{\epsilon\!\!/}}
\newcommand{\vpslash}{ \vslash^\prime}
\newcommand{\gfive}{\gamma_{5}}
\newcommand{\sqrtkq}{\sqrt{m_{K^*} m_Q}}
\newcommand{\mkay}{m_{K^*}}
\newcommand{\gev}{\mbox{GeV}}
\newcommand{\lbar}{\bar{\Lambda}}
\newcommand{\Tr}{\mbox{{\rm Tr}}}

\def\fun#1#2{\lower3.6pt\vbox{\baselineskip0pt\lineskip.9pt
        \ialign{$\mathsurround=0pt#1\hfill##\hfil$\crcr#2\crcr\sim\crcr}}}
\def\mypacs#1#2{\par %
\bgroup
\hsize\columnwidth \parindent0pt
\if@twocolumn\else\leftskip=0.10753\textwidth \rightskip\leftskip\fi
\ifdim\prevdepth=-1000pt \prevdepth0pt\fi
\dimen0=-\prevdepth \advance\dimen0 by20pt\nointerlineskip
\vbox to28pt{\small\vrule height\dimen0 width0pt\relax\ifdraft{#1}\fi\vfill}%
\egroup
\if@twocolumn\vskip1pc\fi
\ifpreprintsty
\penalty10000\vfill
\hbox to\columnwidth{\hfil #2}\newpage
\fi
}
\begin{document}

\draft

\title{Determination of \vts\ from \dkev\ and \bkg\ data \\
via heavy quark symmetry and perturbative QCD}
\author{P.~A.~Grif{fi}n \and M.~Masip
\and M.~McGuigan\footnote{E-mail addresses: \{pgrif{fi}n,
masip, mcguigan\}@phys.ufl.edu}}
\address{Department of Physics, University of Florida,
Gainesville, FL 32611}

\date{\today}

\maketitle

\begin{abstract}

We use heavy quark effective theory (HQET) and perturbative
QCD to study the heavy meson -- light vector meson transitions
involved in $D$ and $B$ decays.
HQET is used to relate the
measured \dkev\ vector and axial-vector form factors at
four-momentum transfer $q^2 = 0$\ to the \bkg\ tensor and
axial-tensor form factors at $q^2 = 16.5$ $\gev^2$.
Perturbative QCD is then used to find matching conditions
for the $B$-meson
form factors at $q^2 = 0$.
A five parameter ``vector
dominance'' type fit of the two HQET form factors,
consisting of single pole, double pole, and subtraction terms,
is used to match the
data at $q^2 = 16.5$ $\gev^2$ to QCD at $q^2 = 0$.
The values
at $q^2 = 0$ are compared with recent data on the exclusive rate for
\bkg\ decay to extract a value for the Cabibbo-Kobayashi-Maskawa matrix
element $\vts\ = 0.035$, with 28\% experimental uncertainty,
and 32\% theoretical uncertainty from higher order QCD effects
and violations of heavy quark symmetry.

\end{abstract}

\mypacs{PACS: 13.20.Fe, 13.20.Jf}{Preprint: UFIFT-HEP-93-25}

\section{Introduction}

Rare $B$ decays are a sensitive probe of flavor changing processes.
In particular, the quark level process $b \rightarrow s \gamma$
occurs mainly via a ``penguin'' type diagram that is sensitive
to the Cabibbo-Kobayashi-Maskawa matrix element $\vts$.
A proper analysis of the penguin requires
a treatment of leading logarithmic strong
interaction corrections\cite{penguin}.  At low energies,
assuming only standard model physics, it
corresponds to the effective Hamiltonian
\begin{mathletters}
\label{hgamma}
\begin{equation}
H_\gamma = \eta \, m_b \, \bar{s} \, \sigma^{\mu\nu} (1 + \gfive )\,
b \,  F_{\mu\nu} + \ldots \ ,
\end{equation}
\begin{equation}
\hbox{where \ \ }
\eta = {G_F \, e \over \sqrt{2} \, 16 \pi^2} \, V_{tb}\,  V^*_{ts}\,
F_2  (m^2_t / m^2_W ) \ .
\end{equation}
\end{mathletters}
The parameter $F_2$ includes the leading logarithmic corrections and
is mildly sensitive to the top quark mass.  It ranges from
$0.59$ for $m_t = 120$ GeV to $0.68$ for $m_t = 210$ GeV.
We will use the value $F_2 = 0.63$, which
corresponds to $m_t = 150$ GeV.

Recently, experimental evidence for this penguin has been found
by the CLEO collaboration via the exclusive decay \bkg\cite{cleo}:
\begin{equation}
{\rm B.R.}(\bkg ) = (4.5 \pm 1.7)\times 10^{-5} \ .
\label{Ethree}
\end{equation}
This is related to the
decay rate via $\Gamma = {\rm B.R.}/ \tau_B$, where we
use the $B$ meson lifetime $\tau_B = (1.29 \pm 0.05)$
ps\cite{pdbook}.
The theoretical expression for the exclusive rate of is given by
\begin{equation}
\Gamma(\bkg ) = {|\eta|^2 m_b^2 \, (A^{(b)}(0) + B^{(b)}(0))^2 \over
 4\pi m_B^3}(m_B^2 - \mkay^2)^3 \ ,
\label{rate}
\end{equation}
where $A^{(b)}(0)$ and $B^{(b)}(0)$ are tensor form factors\cite{bd}
determined by the matrix element
\hbox{$\langle K^{*} | \bar{s}\, \sigma^{\mu\nu}(1+\gfive)
 \, b | B \rangle $} at $q^2 = 0$ and defined
explicitly in section two.
In this paper, we will
study the consistency of the standard model with this data,
by extracting from it a value for \vts. Throughout this paper we
will take
$V_{tb} = 1$
in the
context of the standard model with three generations\cite{pdbook},
although it should be clear from eqn.~(1b) that we are really
calculating the combination $|V_{ts}^* V_{tb}|$.

Unitary of the Cabibbo-Kobayashi-Maskawa mixing with
three generations of quarks places a severe constraint the
mixing angle \vts.  Following the conventions of the
particle data group\cite{pdbook}, $|V_{ts}| \approx |V_{ud}||V_{cb}| =
0.041 \pm 0.007$.  An additional quark generation or the presence
of vectorlike down type quarks (which appear in $SO(10)$ unification)
would invalidate this result.
The penguin itself is quite sensitive to
non-standard model physics, particularly to the exchange
of charged Higgs bosons\cite{penguin,hewett,bbp},
which can easily suppress or enhance the parameter
$F_2$  by a factor
of two.  In principle, this can
be used to place constraints on supersymmetry breaking
parameters\cite{giudice}.

Quark model
calculations of the hadron matrix elements and the
subsequent branching ratio vary by an order
of magnitude\cite{qmodel}, and are therefore unreliable as a
test of the standard model result for \vts.
Instead, we shall apply
heavy quark effective theory (HQET).  This
is a model independent
framework that relates processes where a heavy quark of mass
$m_Q\gg \Lambda_{QCD}$
exchanges momenta smaller than $m_Q$ with the light
degrees of freedom inside the hadron.  HQET makes manifest the
symmetries of the QCD Lagrangian that occur in the infinite
heavy quark limit:  a flavor symmetry which relates the mass
splittings and
decay amplitudes of hadrons with different heavy quark content,
and a spin symmetry which simplifies the mass spectrum and
relates the decay amplitudes of
hadrons with the same heavy quark content,
but with different heavy quark spin.  The
spin symmetry is analogous to the proton spin symmetry
in the hydrogen atom spectrum, which is broken only weakly
by the hyperfine splittings.
(For a review of HQET, see \cite{iwrev,neubert}
and references therein.)
For example, all of the form factors relevant in the semileptonic
decay of a $B$ to a $D$ or $D^*$
(charmed) meson can be given in terms of a single
universal function, to leading order in an expansion
in $\Lambda_{QCD}/m_Q$,
where $m_Q$ corresponds to $b$ and $c$ quarks masses.
This so called Isgur-Wise function depends only on the
relative four velocities of the initial and final
heavy quarks, and is absolutely normalized at zero recoil.

HQET also has implications in heavy hadron to light
hadron (heavy -- light) transitions. The process
$B \rightarrow \pi \ell \nu$ ,{\it i.e.,}
the decay of a heavy pseudoscalar meson ($s^\pi=0^-$) to a light
pseudoscalar meson,
has recently been studied in this context\cite{blnn}.
For the case at hand, (\bkg), we will need to consider the decay of a
heavy pseudo-scalar meson to a light vector meson ($s^\pi=1^-$).
Therefore,
in Section 2 we will
use the tensor formalism\cite{tensor} to establish
the most general form of any transition amplitude
between a heavy pseudoscalar meson and a light vector
meson consistent with HQET.
Again, the strange quark is {\it not} treated as heavy
compared to the QCD scale $\Lambda_{QCD}$.

To extract \vts\ from the data, we need to determine the
combination of tensor form factors $A^{(b)}(0) + B^{(b)}(0)$ which occur
in the \bkg\ decay rate, eqn.~(\ref{rate}).  Here we
apply heavy quark symmetry to relate this combination to the form factors
associated with \dkev\ data\footnote{We use the convention
$\epsilon_{0123}=+1$.},
\begin{eqnarray}
\label{expff}
\langle K^{*}(p^\prime, \epsilon) | \bar{s}\, \gamma^\mu
(1-\gamma_5) \, c | D (p)\rangle  = &&
{2 i \epsilon^{\mu\nu\alpha\beta}\over m_{K^*}m_{D}}
\epsilon^*_\nu \pp_\alpha p_\beta V(q^2)
-(m_D+m_{K^*})\epsilon^{*\mu} A_1(q^2) \nonumber \\
+{A_2(q^2)\over m_D+m_{K^*}} && \epsilon^*\cdot p (p+\pp)^\mu
+{A_3(q^2)\over m_D+m_{K^*}} \epsilon^*\cdot p (p-\pp)^\mu\ ,\
\end{eqnarray}
where $q^2=(p-\pp)^2$.
The form factor $A_3$ leads to unmeasurable contributions
proportional to the electron mass. For $A_1$, $A_2$, and $V$,
we will use the averages over data from
experiments E653\cite{e653} ($D \rightarrow K^* e \nu$) and E691\cite{e691}
($D \rightarrow K^* \mu \nu$) at Fermilab
given in refs.~\cite{casa,amund}:
\begin{equation}
V(0) = 0.95 \pm 0.20 \ , \ \
A_1(0) = 0.48 \pm 0.05 \ , \ \
A_2(0) = 0.27 \pm 0.11 \ .
\label{Ddata}
\end{equation}
These are the values of the form factors given at zero
invariant momentum transfer, $q^2 = 0$, for the \dkev\ system,
extrapolated with a simple pole ansatz for the form factors.
(The results are relatively insensitive to the form of the
ansatz, because the data is taken in the narrow
region $q^2 = 0$ to $q^2 \approx 1\ \gev^2$.)

Heavy quark symmetry will relate the two decay systems in in the following
way:  since to leading order the heavy quark form factors calculated
from the effective theory are independent of the heavy quark mass,
they are functions only of the $K^*$ mass and the dimensionless
parameter
\begin{equation}
w = {v \cdot p^\prime \over  m_{K^*}} \ ,
\label{doubleu}
\end{equation}
where $v$ is the velocity of the heavy meson and $p^\prime$ is the
momentum of the final state $K^*$ meson.  On the
other hand form factors
are given as a function of
\begin{equation}
q^2 = m_Q^2 + m_{K^*}^2 - 2 m_Q m_{K^*} w \ ,
\label{qsquared}
\end{equation}
where $m_Q$ is the mass of the heavy meson.
This means that $q^2 = 0$ for the \bkg\ decay corresponds to
$w \equiv w_B \approx 3.04$, while for the \dkev\ decay, $q^2 = 0$ corresponds
to $w \equiv w_D \approx 1.29$.  We will show in section 2
that the heavy quark flavor and spin
symmetries relate the \dkev\ data given by (\ref{Ddata}) to the
tensor form factors of the \bkg\ decay at $w = w_D$, which corresponds
via eqn.~(\ref{qsquared}) to $q^2 \approx 16.5\ \gev^2$ for the $B$ decay
system.
We must then evolve the $B$ decay form factors down to $q^2 = 0$, or
equivalently, from $w_D$ up to $w_B$.

To determine the interpolating functions that can be used to
evolve the form factors from $w_D = 1.29$ to
$w = 3.04$, which for \bkg\ decay is the kinematical point corresponding to
the emission of an on shell photon, we will use
Perturbative QCD. This is the appropriate description of strong
processes involving the exchange of hard gluons, and corresponds to
heavy -- light transitions with a large value of the
kinematical variable $w$, that is, far away from the ``zero recoil''
point $w=1$.
Therefore, in Section 3 we develop the
Brodsky-Lepage formalism\cite{bl} of perturbative QCD
to calculate the HQET heavy -- light
form factors for ``large'' $w$, to leading order in the heavy
quark mass expansion, and to leading order in $\alpha_s.$
We argue that the method is reliable at
$w= w_B$, (corresponding to $\alpha_s \approx 0.20$),
and use this perturbative calculation to
place ``matching constraints''
on the interpolating form factors.
Furthermore, we use perturbation theory
to determine the leading violations of HQET -- the
order $w\cdot m_{K^*}/m_Q$ corrections.

We will use the \dkev\ data to
constrain the interpolating form factors at $w_D$, and perturbative
QCD to constrain them at $w_B$.  As discussed in section
4, this will enable us to make a five
parameter fit to the two relevant heavy-light form factors
required for the determination of $\vts$.
The parametrization is in terms of single pole, double pole, and a single
subtraction (constant piece), consistent with ``vector dominance''
ideas.

This is a better way of fitting to perturbative
QCD than simply assuming a pole form
for the vector and axial-vector form factors because we require
the parametrization
to match the heavy -- light form factors in the perturbative regime.
This is {\it not} the same as matching the perturbative
result in the limit $w \rightarrow \infty$, with $m_Q$ fixed; in
this limit, perturbative QCD counting rules\cite{counting} do indeed indicate
single pole dominance, but this is due to the $w\cdot m_{K^*}/m_Q$
corrections from the point of view of the HQET.  The HQET limit
corresponds to taking $m_Q \rightarrow\infty$, with $w$ fixed, and
this is the correct way of performing perturbation theory in our
case, since the evolution of the form factors occurs for $w\cdot
m_{K^*}/ m_B < 1$.
In fact, as we shall show in section 2,
assuming a simple pole form for these form factors, with
no subtactions, is inconsistent with heavy quark
symmetry to leading order in heavy quark mass.

As a self consistency check, we use the absolute normalization
of the $B$ decay form factors, obtained by HQET
matching to the $D$ decay data, to
estimate of the B-decay constant $f_B$ to leading order in
QCD perturbation theory.  This check, and further error estimates,
are contained in section 5.

Section 6 is devoted to a discussion on how to
systematically improve these results, and our conclusions.

\section{Heavy -- light form factors}

We shall first apply the tensor method\cite{tensor}
to the case of a heavy pseudoscalar meson $M_Q(v)$
decaying into the light vector meson $K^*(\epsilon,\pp)$.
This method is usually applied to the derivation of heavy to heavy meson
form factors\cite{iwrev,neubert}, and recently has been used
for the case of the heavy meson to light meson transition
$B \rightarrow \pi \ell \nu$\cite{blnn}.  We will determine
the most general form of heavy psuedoscalar to light vector
form factors consistent with Lorentz invariance
and heavy quark spin symmetry.

It is useful to consider the matrix element of the
general operator $\bar{s} \Gamma Q$, where $\Gamma$ is an arbitrary
product of gamma matrices, and $Q$ is a heavy quark Dirac field.  To leading
order in the HQET,
$Q$ is replaced by its projection onto the quark (versus antiquark)
field via the replacement $Q \rightarrow h^{(Q)}_v$, where
$h^{(Q)}_v (x) = \exp {(i m_Q v\cdot x)} {1 + \vslash \over 2} Q (x)$,
and $v$ is the heavy meson velocity obeying $v^2 = 1$.
The relevant matrix element takes the form
\begin{equation}
\langle K^{*}(p^\prime, \epsilon) |
\bar{s}\, \Gamma \, h^{(Q)}_v | M_Q (v)\rangle
= \sqrtkq \, {\rm Tr} \left\{\Theta (v, \pp, \epsilon^* ) \, \Gamma\,
{1 + \vslash \over 2} \gfive \right\} \ ,
\label{one}
\end{equation}
where $p^\prime_\mu  = \mkay \vp_\mu $ is the $K^*$ momentum,
and $\epsilon$ is the
$K^*$ polarization vector which satisfies $\epsilon^* \cdot \pp = 0$.
The ${1 +\vslash \over 2}$ in
Eq.\ (\ref{one}) explicitly projects the matrix element
onto the heavy quark components of the heavy quark spinor,
since in the infinite quark mass limit heavy antiquarks are not
produced, and the $\gfive$ describes the pseudoscalar nature of the
heavy meson.
A simple way to understand eqn.~(\ref{one}) is by parametrizing the
initial heavy meson as $| h^{(Q)}_v \gfive \ell \rangle$, where
$\ell$ describes all of the light degrees of freedom in the heavy
meson, and using the heavy quark propagator\cite{iwrev} $\langle
h^{(Q)}_v {\bar{h}}^{(Q)}_v \rangle = (1 + \vslash)/2$ .  The matrix
function $\Theta$ encapsulates all of our ignorance of the dynamics of
the light degrees of freedom of the heavy meson and the $K^*$
meson.   It is explicitly independent of the heavy quark mass to
leading order in HQET.  The $\sqrt{m_Q}$ in eqn.~(\ref{one}) comes from
the usual normalization of the heavy meson with respect to its energy.

More specifically, the scalar matrix $\Theta$ must
be proportional to the polarization of the $K^*$ meson:  $\Theta =
{\cal M}_1 \esslash + {\cal M}_2 v\cdot \epsilon^*$, where the
${\cal M}_i$ are matrix functions of $\mkay$ and $w$ defined
in eqn.~(\ref{doubleu}).  Furthermore, using the fact that
${1 + \vslash\over 2}\vslash = {1 +\vslash\over 2}$,
the only possible matrix structure of each
${\cal M}_i$ within the trace is $\theta + \vpslash \theta^\prime $,
where the $\theta$ and $\theta^\prime$ are real functions of $v\cdot \pp$.
Therefore, one can parametrize $\Theta$ in terms of only four
linearly independent form factors, $\theta_i (\mkay, w )$,
\begin{equation}
\Theta = ( \theta_1 + \vpslash \theta_2 )\esslash +
(\theta_3 + \vpslash \theta_4) v \cdot \epsilon^* \ .
\label{two}
\end{equation}
We will refer to Eqs.\ (\ref{one},\ref{two}) as the heavy -- light
matrix elements for the decay of a heavy pseudoscalar into a light
vector.  This is the relativistic generalization (where the
heavy quark is not necessarily in its rest frame)
of the heavy -- light matrix elements obtained in Ref.\cite{iwone}

Since the $b$ and $c$ quarks are not infinitely heavy, the relation
between operators in QCD and in the heavy quark effective theory
is non-trivial.  In the leading logarithmic approximation,
the relation between the QCD currents and
heavy quark effective theory currents discussed above is
$\bar{s} \Gamma Q = C_Q(\mu) \bar{s} \Gamma h_v^{(Q)}$,
where the coefficient functions are\cite{lla}
$C_Q(\mu) = \left[ \alpha (m_Q) / \alpha(\mu)
\right]^{-6/(33 - 2 N)} $,
$N$ denotes the number of flavors below the $m_Q$ scale,
and $\mu$ is the infrared subtraction point.  It is useful to
define $\theta^{(Q)}_i = C_Q \theta_i$.  The $\theta^{(Q)}_i$ are
subtraction point independent and satisfy
\begin{equation}
\theta^{(b)}_i = C_{cb} \, \theta^{(c)}_i \ ,
\  \
C_{cb} = \left[ {\alpha (m_b) \over \alpha(m_c)}
\right]^{-6/25} \ .
\label{llog}
\end{equation}
We use $\alpha_s(m_b)/\alpha_s(m_c)
=  [\ln {(m_D^2 / \Lambda_{QCD})} / \ln {(m_{B}^2
/ \Lambda_{QCD})}]$, which for $\Lambda_{QCD} = 300$ MeV gives
$C_{cb} = 1.114$.

Eqs.\ (\ref{one}, \ref{two}) place strong constraints
between the form factors corresponding to different matrix elements.
For $\Gamma = \sigma^{\mu\nu}$, the matrix element is
given by
\begin{equation}
\langle K^{*}(p^\prime, \epsilon) | \bar{s}\, \sigma^{\mu\nu}
 \, Q | M_Q (p)\rangle =
 \epsilon^{\mu\nu\alpha\beta}\left(
 A^{(Q)} \epsilon^*_\alpha p_\beta +
 B^{(Q)} \epsilon^*_\alpha \pp_\beta +
 C^{(Q)} \epsilon^* \cdot p \, p_\alpha \pp_\beta \right) \ .
\end{equation}
Using the identity $\sigma^{\mu\nu}\gfive = -{1\over 2}i
\epsilon^{\mu\nu\alpha\beta} \sigma_{\alpha\beta}$ one can immediately
write the form factors for the $\Gamma = \sigma^{\mu\nu} \gfive$
matrix element,
\begin{eqnarray}
\langle K^{*}(p^\prime, \epsilon) | \bar{s}\, \sigma^{\mu\nu}\gfive
 \, Q | M_Q (p)\rangle =
i \bigg[ && A( \epsilon^{*\mu} p^\nu - \epsilon^{*\nu} p^\mu )
+ B (\epsilon^{*\mu} \pp^\nu - \epsilon^{*\nu} \pp^\mu ) \nonumber \\
&& \mbox{} + C \epsilon^* \cdot p \, (p^\mu \pp^\nu - p^\nu \pp^\mu )
\bigg] \ .
\end{eqnarray}
Evaluating these matrix elements using
Eqs.\ (\ref{one},\ref{two}) gives relations between the ``tensor''
form factors $A,B,C$, and the heavy -- light form factors $\theta_i$,
\begin{equation}
A^{(Q)} = -2 \sqrt{\mkay\over m_Q}\theta^{(Q)}_1 \ , \ \
B^{(Q)} = 2 \sqrt{m_Q\over \mkay}\theta^{(Q)}_2 \ , \ \
C^{(Q)} = {2 \, \theta^{(Q)}_4 \over m_Q^{3/2} \mkay^{1/2} }\ .
\label{ABC}
\end{equation}
The vector and axial-vector matrix elements can be parametrized as
\begin{mathletters}
\label{seven}
\begin{equation}
\langle K^{*}(p^\prime, \epsilon) | \bar{s}\, \gamma^\mu
 \, Q | M_Q (p)\rangle =
i D^{(Q)} \epsilon^{\mu\nu\alpha\beta} p_\nu \epsilon^*_\alpha
\pp_\beta \ ,
\end{equation}
\begin{equation}
\langle K^{*}(p^\prime, \epsilon) | \bar{s}\, \gamma^\mu\gfive
 \, Q | M_Q (p)\rangle =
E^{(Q)} \epsilon^{*\mu} + F^{(Q)} \epsilon^* \cdot p p^\mu
+ G^{(Q)} \epsilon^* \cdot p \pp^\mu \ .
\end{equation}
\end{mathletters}
Evaluating these matrix elements using the heavy -- light
parameterization yields
\begin{mathletters}
\label{detheta}
\begin{equation}
D^{(Q)} = {-2\, \theta^{(Q)}_2 \over \sqrt{\mkay m_Q}} \ , \ \
E^{(Q)} = 2\sqrt{\mkay m_Q} \theta^{(Q)}_1 -
{2\, p\cdot \pp \over \sqrt{\mkay m_Q}} \theta^{(Q)}_2 \ ,
\end{equation}
\begin{equation}
F^{(Q)} = -2 \sqrt{\mkay m_Q} { \theta^{(Q)}_3 \over m_Q^2} \ , \ \
G^{(Q)} = {2( \theta^{(Q)}_2 + \theta^{(Q)}_4)\over
\sqrt{\mkay m_Q}} \ .
\end{equation}
\end{mathletters}
Since the seven form factors for the matrix elements are
given in terms of only four heavy -- light form factors, there are
three relations relations between them for each heavy meson $M_Q$:
\begin{equation}
A =  {-E + (p\cdot\pp) D \over m_{Q}}\ , \ \
B = - m_{Q} D \ , \ \
C = { G + D \over m_{Q}} \ .
\label{nine}
\end{equation}
These relations explicitly relate the tensor form factors to the
vector and axial-vector form factors for a given heavy meson $M_Q$
transition to $K^*$.  They represent the explicit realization of the
heavy quark spin symmetry for the heavy pseudoscalar meson decay to
a light vector.  Equivalent relations, determined via
analysis in the heavy quark rest frame, are found in
refs.~\cite{bd,iwone}.  This confirms our analysis which
led to eqns.~(\ref{one},\ref{two}).

The relationships between the heavy -- light form factors
at the kinematical point $w_D = (m_D^2+m_{K^*}^2)/ 2 m_Dm_{K^*} \approx
1.29$ and the experimentally measured \dkev\ form factors are
\begin{mathletters}
\label{twelve}
\begin{eqnarray}
\theta^{(c)}_1 (w_D) && =  {m_D + \mkay \over 2 \sqrt{m_D \mkay}}
\left[ A_1(0) - {\mkay^2 + m_D^2 \over (m_D + \mkay )^2} V(0) \right]\ , \\
\theta^{(c)}_2 (w_D) && =  -{\sqrt{ \mkay m_D} \over \mkay + m_D} V(0) \ , \\
\theta^{(c)}_3 (w_D) && - {m_D\over \mkay} \theta^{(c)}_4 (w_D)
= \sqrt{m_D\over \mkay}{m_D\over m_D+\mkay }[A_2(0)-V(0)]\ , \\
\theta^{(c)}_3 (w_D) && + {m_D\over \mkay} \theta^{(c)}_4 (w_D)
= \sqrt{m_D\over \mkay}{m_D\over m_D+\mkay }[A_3(0)+V(0)]\ .
\end{eqnarray}
\end{mathletters}
{}From the data, the values of $\theta_1$, $\theta_2$
and the combination $\theta_3-(m_D/\mkay)\theta_4$ at $w_D$ can be
extracted,
\begin{mathletters}
\label{thvalues}
\begin{equation}
\theta^{(c)}_1(w_D)=-(0.06\pm 0.13)\ , \ \
\theta^{(c)}_2(w_D)=-(0.44\pm 0.09)\ ,
\end{equation}
\begin{equation}
\theta^{(c)}_3(w_D)-{m_D\over \mkay} \theta^{(c)}_4(w_D)
= -(0.66\pm 0.22)\ ,
\end{equation}
\end{mathletters}
where we have quoted experimental uncertainty.

Treatment of the strange quark as heavy in the context of HQET
has been considered in the literature\cite{strange}.  Eqns.~(\ref{thvalues})
indicate that $\Lambda_{QCD}/m_s$ corrections are large.
In an expansion in terms
of the strange quark mass, the trace formula eqn.~(\ref{one}) for the
heavy -- light form factor is modified by addition of $\esslash (1 +
\vpslash )/2$, to project onto the ``heavy'' strange quark (verses
antiquark).  This exercise gives in the standard result for the
Isgur-Wise function $\xi$.  Using the heavy -- light parametrization,
one finds that $\xi = \theta_1 - \theta_2$, and further that
$\theta_1 + \theta_2$,
$\theta_3$, and $\theta_4$ each vanish to leading order in the
$1/m_s$ expansion.  Clearly, the vanishing relations are strongly
violated by the data given above.

Heavy quark flavor symmetry relates the heavy -- light
form factors for $D$ decay to those for $B$ decay, as given in
equation (\ref{llog}).  The data determines $\theta_1$ and $\theta_2$,
which in turn give the tensor form factors required for the
extraction of \vts\ via eqn.~(\ref{ABC}).  We are then left to determine
the running of the form factors from $w_D$ to $w_B$.  This is
a dynamical question, and the kinematical heavy quark symmetry
cannot give us the answer.

There are two
``straightforward'' avenues of approach to this problem,
QCD sum rules\cite{cz}, and perturbative QCD\cite{bl}.  In
the next section, we will discuss the later approach.  We note
that in the literature, a third approach is discussed, which is to
make an outright guess
as to the $q^2$ dependence of the form factors, following
vector dominance ideas.  To see the danger in this, consider assuming the
standard single pole vector dominance ansatz for the vector and axial
vector form factors $V(q)$ and $A_1(q)$ for the $B$ meson decay.
This means that $V$ and $A_1$ are of the form
$m^2_T/(q^2 - m^2_T)$,
where $m_T$ a vector or axial vector threshold mass above the $B$ meson
mass.  Via eqns.~(\ref{doubleu},\ref{qsquared}), in terms of the
more relevant HQET variable $w$, single pole dominance is of the
form $1/w$, plus order $\Lambda_{QCD} / m_Q$ corrections.  However,
now that we have correctly parametrized the heavy -- light form factors,
it becomes clear that
this form is incompatible with leading order HQET.  Since $V \propto
\theta_2$, and $A_1 \propto (\theta_1 - w \theta_2)$, $A_1$ must have
a ``large'' subtraction if the vector form factor obeys single
pole dominance.

\section{\bkg\ form factors at large $w$ from perturbative QCD}

In the previous section we have used HQET to relate the form factors
in $D \rightarrow K^{*} e \nu $ with those in
$B \rightarrow K^{*} \gamma$ at $w_D=1.29$. However the physical value
of $w$ for $B \rightarrow K^{*} \gamma $ that corresponds to an
on shell photon is $w_B=3.05$. This value for the $B$ decay is
in the regime of large $w$ far from zero recoil. How does
one estimate the behavior of the form factors in this regime?
First consider the treatment of heavy meson to heavy meson
transitions by heavy quark symmetry. All form factors are determined
from the Isgur-Wise function $\xi (w) $ and heavy quark symmetry
fixes the value $\xi (w) = 1$. For large enough values of $w$
the Brodsky-Lepage method of determining form factors from
perturbative QCD can be invoked. To leading order, the Brodsky-
Lepage formalism essentially includes the exchange of a single
hard gluon between the initial or final heavy quark and its
spectator quark.
Assumptions about the meson wave functions, motivated by QCD sum rules
and experiment\cite{cz} are made to fix the soft behavior of the
amplitude.
This method determines\cite{kk} the behavior of the Isgur-Wise function
at large $w$.  For $m_Q/\Lambda_{QCD} > w \gg 1$,
$\xi (w) \sim {\alpha_s(m_Q) \over {w^2}}$.  As discussed
in the previous section, this corresponds to a dipole
in the sense of a vector dominance model.
For the true asymptotic regime of $w > m_Q/\Lambda_{QCD}$,
$\zeta(w) \sim {\alpha_s(Q)\over {w}}$, where $Q^2 = -q^2$.  This is
in accordance with QCD dimensional-counting rules\cite{counting}.

We shall essentially follow the same procedure in discussing heavy
meson to light meson transitions.
In this case of course we have
$\theta_1, \theta_2, \theta_3$ and $\theta_4$ instead of the single
Isgur-Wise function. Although we don't have the convenient normalization
at $w=1$, we can use instead the $D \rightarrow K^{*} e \nu$ data to
determine
the thetas at $w=1.29$. The fact that the kinetic point corresponding
to $B \rightarrow K^{*} \gamma$ at $w=3.05$ is far from zero recoil
is a virtue; it is in just the right regime for
both perturbative QCD and heavy quark symmetry to be valid. That is,
if $w$ is close to $1$, perturbative QCD is not valid, while if
$w > m_Q/\Lambda_{QCD}$, then $w\cdot \Lambda_{QCD}/m_Q$ corrections
dominate the HQET.

We now briefly describe standard aspects of the Brodsky-Lepage
formalism as it applies to the case at hand.
The soft nonperturbative part of the physics is encapsulated in
quark distribution amplitudes $\phi(x,P,Q)$, for each
meson, which denote the fraction $0 < x < 1$ of momentum $P$
carried by the valence quark of the meson. To leading order in
the calculation,
the antiquark carries momentum $(1-x)$; it can be argued that
gluon and non-valence quark corrections are small for large momentum
exchange via the QCD dimensional-counting rules.  The parameter
$Q$ denotes the subtraction point at which the distribution
amplitudes are evaluated.  Given a distribution amplitude $\phi$ at
$Q_0$, $\phi (Q)$, for $Q > Q_0$ can be determined\cite{bl}.
The dependence is
logarithmic in $Q$, and $\phi$ asymptotically approaches $\phi \propto
x (1-x)$.  We shall initially consider two distribution amplitudes
for the $K^*$ meson, as shown in fig.~1,
\begin{figure}[t]
\begin{center}
\leavevmode
\epsfxsize=4truein
\epsffile{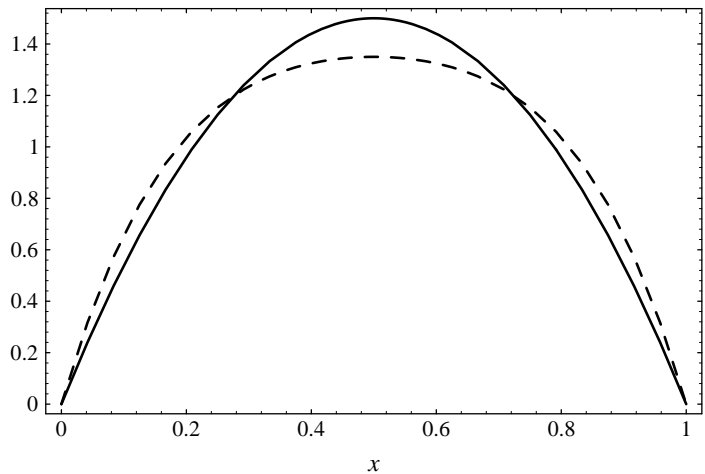}
\end{center}
\caption{Two choices for the $K^*$ distribution
amplitude.  The solid curve is $\bar{\phi}_1 (x)$, and the dashed curve
is $\bar{\phi}_{czz} (x)$, as defined in the text.  Both are normalized to
unit area.}
\end{figure}
\begin{mathletters}
\label{das}
\begin{equation}
\bar{\phi}_{1} = 6 x ( 1 - x ) \ .
\end{equation}
\begin{equation}
\bar{\phi}_{czz} = - 12 x^2 (1-x)^2 + {42\over 5} x (1 - x) \ .
\end{equation}
\end{mathletters}
The second of these is the Chernyak-Zhitnitsky-Zhitnitsky\cite{czz}
distribution amplitude for the $K^*$ at $Q^2 = 1.5\ \gev^2$.  The
integrals over $x$ for the functions
are both normalized.  However, the correct
normalization is given by $\phi = f_{K^*} \bar{\phi}/2\sqrt{6}$, where
$f_{K^*}$ is the meson decay constant (With this normalization,
the pion decay constant $f_{\pi} = 131$ MeV\cite{pdbook}).

For the momentum fraction of the heavy $b$
quark inside the initial $B$, we use the peaking
approximation.
The distribution amplitude for heavy quarks must have a large
maximum near the end point $x= 1- {\lbar/m_Q}$, and
width  $\lbar/m_Q$, where $m_Q = m_q + \lbar$.  We
use\cite{neubert} $\lbar = 0.5\ \gev$ as the typical amount of
mass in the heavy meson due to the light degrees of freedom
and make the approximation
\begin{equation}
\label{bdist}
\phi_Q(x) = {1 \over 2 \sqrt{6} }
f_Q \delta(1-x-{\lbar\over m_Q}) \ .
\end{equation}
The error induced by this approximation is of order $\lbar/m_Q$, which
we are systematically neglecting in this paper.

The initial and final states are given by convoluting the distribution
amplitudes with on-shell spinors of the quark and antiquark,
\begin{equation}
\Psi = \phi \sum_{\lambda,\lambda^\prime} f_{\lambda\lambda^\prime}\,
u_{\lambda} (x P) \, \bar{v}_{\lambda^\prime}( (1-x) P) \ ,
\end{equation}
where $\lambda$ and $\lambda^\prime$ denote sums over helicities.
In the context of the peaking approximation,
$(\Pslash - m_Q) u_{\lambda} (xP) = 0$,
and $(\Pslash + m_Q ) v_{\lambda^\prime} = 0$.
By applying these equations to the
helicity sums, we find
\begin{equation}
\Psi_Q = {\phi_Q (x)\over\sqrt{2}}
(m_Q + \Pslash )\gfive \ .
\end{equation}
For the $K^*$, we replace the $\gfive$ with $\epslash$, and
let $m_Q \rightarrow m_S$.  The last substitution is strictly
not rigorous, because we are not using a peaking approximation for the
$K^*$ distribution amplitude.
This will induce errors of order $m_{K^*}/M_B$ in
the final results.  Required for the calculation is the
conjugate of the state, $\bar{\Psi}_{K^*}= \gamma^0 \Psi_{K^*}^{\dagger}
\gamma^0$,
\begin{equation}
\bar{\Psi}_{K^*} = - {\phi_{K^*} (x)\over\sqrt{2}}
\esslash(m_{K^*} + \Pslash )\ .
\end{equation}

Our calculation will include the exchange of a single hard gluon between
the initial heavy or $s$ quark and its spectator quark.
The two diagrams are shown in figure 2(a,b),
representing respectively the single gluon exchange between the
heavy or $s$ quark and its spectator. Here $\Gamma$ represents
the operator responsible for the heavy to light
transition, as in the previous section.
\begin{figure}[t]
\begin{center}
\leavevmode
\epsfxsize=6truein
\epsffile{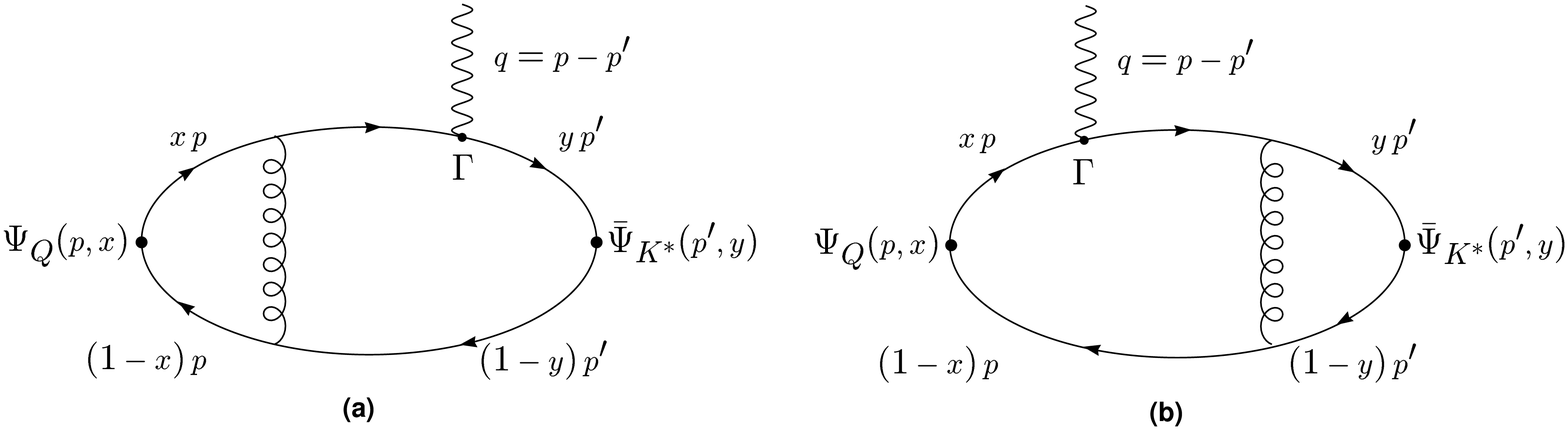}
\end{center}
\caption{The two Born amplitudes
contributing to $M_Q \rightarrow K^*$ decay.  The symbol $\Gamma$
denotes an arbitrary Dirac matrix associated with the decay.}
\end{figure}
These diagrams
yield the contribution to the hard scattering amplitude $T_{\Gamma}$
from perturbative QCD.
We use the identity $T^a T^a = {4\over 3}I_c$ where
$I_c$ is the 3x3 identity matrix of color space.
The Brodsky-Lepage formalism tells us
that the full amplitude at large $w$ is given by
\begin{equation}
<K^* | \bar{s}\Gamma Q |M_Q> =
\int dx dy \Tr \bar{\Psi}_{K^*}(y) T_{\Gamma}(y,x)\Psi_Q(x)$$
\end{equation}

Both diagrams are infared (IR) divergent when the momentum flowing
through the gluon line vanishes.  Fig. (2b) also has an IR divergence
associated with the strange quark going on-shell for small but finite
gluon momentum transfer.  These IR divergences are fictitious\cite{bl}
and irrelevant at large recoil.
The region of $y$ integration associated with the IR problem
corresponds to the Drell-Yan-West\cite{dyw} end point region,
which is assumed to be cut off by a Sudakov form factor.
Physically, the mutually canceling effects of multi-gluon exchange
cut off this region, due to the fact that the color singlet meson
decouples from soft gluons in the IR limit.

Following the cited literature, we apply a transverse momentum
cutoff to suppress IR effects.  In the rest frame of the
heavy quark, we have the following picture.  The
momentum of the $K^*$ is given by $\mkay (\cosh \theta , 0 , 0, \sinh
\theta)$.  In this frame, the variable $w = \cosh \theta$, and the
spacelike gluon momentum is $k_3 = (1-y)\mkay \sinh \theta$.  This
is the transverse momentum of the gluon, associated with transverse
momentum of $p  - p^\prime$.  Large transverse momentum means that
this is larger than the typical momentum associated with the light
degrees of freedom, $\lbar$.
Therefore we apply the cutoff
\begin{equation}
1-y > {\lbar \over m_{K^*} \sqrt{w^2-1}}.
\label{cutoff}
\end{equation}


The calculation is done in Feynman gauge.  To extract the heavy -- light
form factors, an expansion of the perturbative expression in powers
of $1/m_Q$ must be done.  The leading terms for the
gluon, strange quark, and heavy quark
denominators ($k^2 - m^2$ in each case), in the peaking
approximation, are given by
\begin{mathletters}
\label{denominators}
\begin{eqnarray}
D_g & = & (1 - y)^2\mkay^2 + \lbar^2 - 2\lbar \mkay (1-y) w \ , \\
D_s & = &\mkay^2 - m_s^2 + \lbar^2 - 2\lbar \mkay w
\equiv m_{K^{*}}\Delta_s \ , \\
D_Q & = & 2 m_Q ( \lbar - (1 - y)\mkay w) \equiv m_Q \Delta_Q
\end{eqnarray}
\end{mathletters}
By comparing the perturbative expression to the heavy -- light
parametrization (\ref{one}), we find
\begin{mathletters}
\label{pertthetas}
\begin{eqnarray}
\theta_1 & = & \kappa (I_Q - {{\lbar - m_s}\over \mkay}I_S) \ , \\
\theta_2 & = & -\kappa(I_Q + I_S) \ , \\
\theta_3 & = & -\kappa {2\lbar\over \mkay} I_s \ , \\
\theta_4 & = & 0 \ ,
\end{eqnarray}
\end{mathletters}
to leading order in the $1/m_Q$ expansion.  The constant $\kappa$ is
\begin{equation}
\kappa = f_{K^*}f_Q {\sqrt{m_{K^*}m_Q} \over m_{K^*}^3}
          {4\pi \alpha_s \over 9} \ .
\label{Ekappa}
\end{equation}
{}From general heavy quark
symmetry arguments\cite{neubert} and the perturbative calculation
done here, one finds that $f_Q\sim 1/\sqrt{m_Q}$ for large mass $m_Q$,
and therefore $\kappa$ is finite and non-vanishing in the infinite
heavy quark limit.  The scale of $\alpha_s$ is set by the $M_Q$ meson
mass.  We will use for the $B$ meson $\alpha_s = 0.20$.
The functions $I_Q$ and $I_s$ denote the integrals
over the momentum fraction $y$ from figure (2a) and (2b) respectively,
\begin{mathletters}
\begin{eqnarray}
I_Q & = &\int_{0}^{1-\epsilon} dy \bar{\phi}^\dagger_{K^*} (y)
{\mkay^3 \over \Delta_Q D_g} \ , \\
I_s & = & \int_{0}^{1-\epsilon} dy  \bar{\phi}^\dagger_{K^*} (y)
{\mkay^3 \over \Delta_s D_g} \ ,
\end{eqnarray}
\end{mathletters}
where $\epsilon$ denotes the IR cutoff given by eqn.~(\ref{cutoff}).
We provide the exact integrals for arbitrary wave function and cutoff
in appendix A. For the distribution amplitudes $\phi_1$ and
$\phi_{czz}$ defined above, they are plotted in fig.~3.
\begin{figure}[t]
\begin{center}
\leavevmode
\epsfxsize=4truein
\epsffile{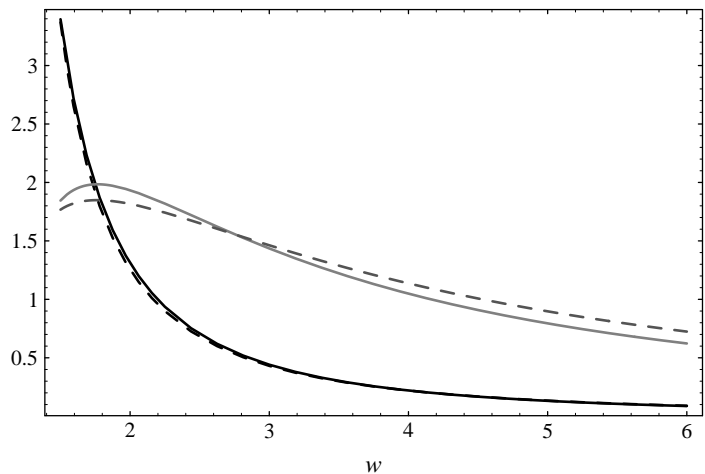}
\end{center}
\caption{The two integrals $F_S$ (black curve) and $F_Q$ (gray curve) as
defined in the text.  The solid (dashed) curves are for
the $\phi_1$ ($\phi_{czz}$) $K^*$ distribution amplitude.
}
\end{figure}
{}From these figures, one can observe that the Drell-Yan-West region
is turning over $I_Q$ at $w\sim 2$; that is, the loss of integral
due to the IR cutoff is ``beating'' the IR divergence around this region.
The integral $I_Q$ turns over at a smaller value of $w$ because of the
additional strange quark pole.  For the remainder of this paper, we will
study only the $\phi_1$ case.

To study the region (in $w$) of validity of this
calculation, one can compare the asymptotic expansion, in powers
of $1/w$, to the full result for $\theta_1$ and $\theta_2$. To leading
orders in this expansion,
\begin{mathletters}
\label{thasymp}
\begin{eqnarray}
\theta_1 (w) &\sim & {\ln w \over w^2} \ , \\
\theta_2 (w) &\sim & {\ln w \over w^2} - {0.23\over w^2} \ .
\end{eqnarray}
\end{mathletters}
\begin{figure}[t]
\begin{center}
\leavevmode
\epsfxsize=4truein
\epsffile{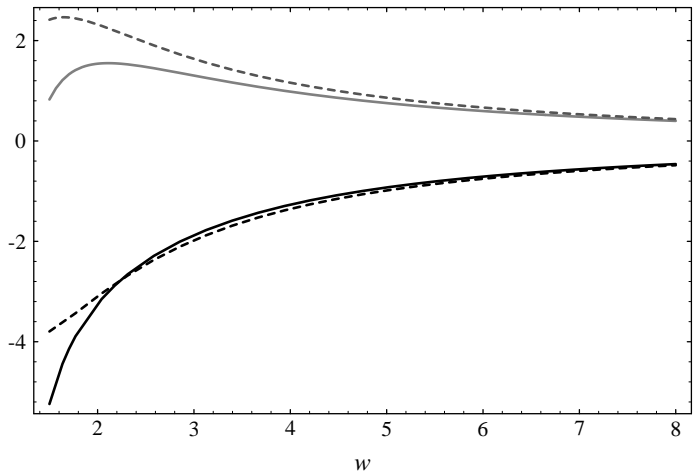}
\end{center}
\caption{
The un-normalized heavy-light form factors $\theta_1/\kappa$
(gray) and $\theta_2/\kappa$ (black), with their asymptotic behavior (dashed).
}
\label{figure4}
\end{figure}
We note that $\theta_1$ also possesses a ${1/ w^2}$ term but in our case it
is numerically small.
In fig.~4, the full and asymptotic (to order $1/w^2$) values of
$\theta_1/\kappa$ and $\theta_2/\kappa$ are plotted, for the
$\phi_1$ distribution amplitude.
For $w < 2 $, less than two thirds of the integration region over
$y$ is included by the Drell-Yan-West cutoff, indicating a large soft
cutoff dependence to the amplitude. Our perturbative QCD result does
not give a reliable estimate of the form factor in this regime.
For $w = 3$, the Drell-Yan-West
region cuts off about $18\%$ of the momentum fraction integral, and
the match between the full and asymptotic functions is reasonable,
as shown in fig.~4.
For $w > 6$ the next to leading order $w\  \mkay / m_Q$ corrections
begin to dominate, and the leading order $\theta_i$ do not meaningfully
describe the amplitude.  We will discuss the effect of the next
to leading order corrections in sec.~5.

Note that the parametric form of the thetas given by
eqn.~(\ref{pertthetas}) has the correct heavy -- heavy limit.
As $m_s$ is taken to infinity,
the integrals $I_Q$ and $I_s$ remain finite and non-vanishing,
and $\xi = \theta_1 - \theta_2$ is the leading contribution.


\section{Interpolating functions and determination of \vts}

For $w$ less than or equal to about 2,
interpolating functions for the heavy -- light form
factors are definitely required.  Our strategy is
to obtain such an interpolating function that is
consistent with the data for
$D \rightarrow K^* e \nu $ at $w=1.29$,
matches the QCD calculation at $w=3.05$, and is consistent with
vector dominance ideas.  For the calculation of $V_{ts}$, we need to
find interpolating functions for $\theta_1$ and $\theta_2$.
As input to determine their forms, we use the two data points
at $w = w_D$, and the three ratios
\begin{equation}
r_1 = {\theta_1^\prime (w_B) \over\theta_1 (w_B)}, \ \
r_2 = {\theta_2^\prime (w_B) \over\theta_2 (w_B)}, \ \
r_3 = {\theta_1 (w_B) \over \theta_2 (w_B)} \ .
\label{ratios}
\end{equation}
These three ratios are determined by perturbative QCD
in the context of the Brodsky - Lapage formalism.
For the $\phi_1$ distribution amplitude, $r_1 = 0.2776$,
$r_2 = 0.4403$, and $r_3 = - 0.6991$.
These
values can be systematically improved by higher order (in
$\alpha_s$) calculations\cite{field}.  As discussed in the
next section, we have reason to believe that the $\alpha_s$
contribution is a good approximation to the full result.
We will consider a five parameter fit to the
two form factors.  Vector dominance ideas, that the
amplitude is dominated by intermediate $b \bar{s}$ resonances,
dictate that reasonable forms for the interpolating
form factors are $1/w$ (single pole), $1/w^2$ (dipole), or a
constant (subtraction).  A subtraction is not considered for
$\theta_2$ because the physical axial-vector form factor
$E \sim (\theta_1 - w \theta_2) $, so that a constant term
in $\theta_2$ would correspond to $E \sim w + \ldots$ which is
``inconsistent'' with vector dominance.

The interpolating functions used are
\begin{mathletters}
\label{thetafit}
\begin{eqnarray}
\theta_1 & = & a \left( {c\over w^2} + {d\over w} \right) + b \ , \\
\theta_2 & = & a \left( {e\over w^2} + {f\over w} \right) \ .
\end{eqnarray}
\end{mathletters}
\begin{figure}[t]
\begin{center}
\leavevmode
\epsfxsize=4truein
\epsffile{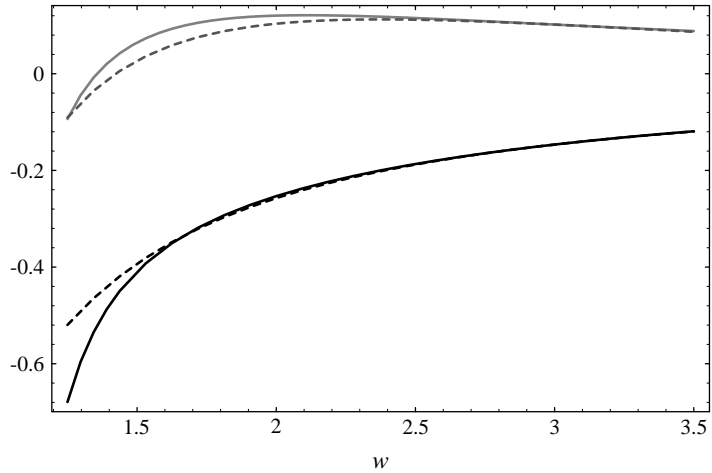}
\end{center}
\caption{
The normalized heavy-light form factors $\theta_1$
(gray) and $\theta_2$ (black), with their
parametrizations with respect to the $K^*$ data (dashed).
}
\label{figure5}
\end{figure}
The theoretical constraints, that the slopes of $\theta_1, \theta_2$
and the ration $\theta_1/\theta_2$ at $w_B$ match the perturbative result
give
\begin{mathletters}
\begin{eqnarray}
c & = & - (r_3 - b/a)w_B^2 + r_1 r_3 w_B^3 \ , \\
d & = & 2 w_B (r_3 - b/a) - r_1 r_3 w_B^2 \ , \\
e & = & - w_B^2 + r_2 w_B^3 \ , \\
f & = & 2 w_B - r_2 w_B^2 \ ,
\end{eqnarray}
\end{mathletters}
so that experiment fixes $a$ and $b$.  The fit to experiment is shown
in fig.~5.  The dashed lines refer to the parametrized fit, and solid
lines refer to the full IR sensitive functions.  By construction,
the fit and original function match very well at $w = w_B$.
(Actually, the full perturbative functions
extrapolate to the data reasonably well without any fits.  This
is another indication that the perturbative calculation is
sensible.).
The propagation of experimental uncertainty for $V$ and $A_1$
from $w_D$ given by eqn.~(5) to $w_B$ via our interpolating
functions is shown in fig.~6.  The dashed lines denote the boundaries
of one standard deviation envelopes about the mean values.
\begin{figure}[t]
\begin{center}
\leavevmode
\epsfxsize=4truein
\epsffile{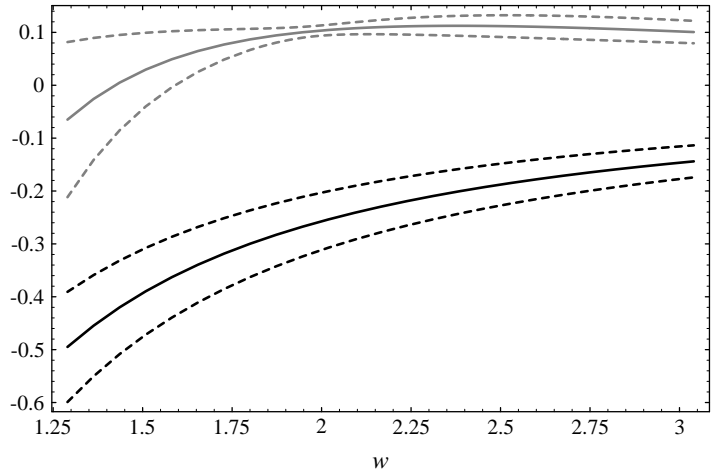}
\end{center}
\caption{
The parametrized heavy-light form factors $\theta_1$
(gray) and $\theta_2$ (black), with one standard deviation
of uncertainty (with respect to the $D \rightarrow K^* \ell
\nu$ data $V(0)$ and $A1(0)$) denoted by the dashed curves.
}
\label{figure6}
\end{figure}

The relevant function for the calculation of $\vts$ is the combination
of tensor form factors $|A^{(b)} + B^{(b)}|$, as discussed in sec.~1,
and defined in sec.~2.  From eqns.~(\ref{ABC}) and the
interpolating function displayed in fig.~6, we find
\begin{equation}
|A^{(b)} + B^{(b)}| (q^2 = 0) = 0.78 \pm 0.16 \ (21\%) \ .
\label{AplusB}
\end{equation}
By combining the CLEO data for \bkg,
the $B$ lifetime, and the theoretical decay rate, given in sec.~1,
with this result, we find
\begin{equation}
\vts = 0.035 \pm 0.010 \ (28\%) \ .
\label{vts}
\end{equation}
Here we have quoted only the experimental uncertainty,
about $21\%$ from \dkev\ data, and $19\%$ from \bkg\ data,
which in turn are added in quadrature.
The theoretical uncertainties are discussed in the
next section.

This result compares quite well with the standard
model value $\vts = .041 \pm .007\ (17\%\ {\rm total\ uncertainty})$
extracted from $V_{cb}$ and the unitarity
of the CKM matrix (discussed in sec.~1).
Therefore our analysis shows that the \bkg\ data agrees
with the standard model, and should be
used to place constraints on non-standard model physics.

\section{Theoretical uncertainties and $f_B$}

Now consider theoretical corrections to our result for
\vts\ given by eqn.~(\ref{vts}).  We consider
corrections to the heavy quark symmetry relations,
to the $\alpha_s (m_B)$ matching conditions,
and uncertainty in the top quark mass.

The corrections to heavy quark symmetry relations are
due to physics which violates the assumption
that in the rest frame of the heavy meson, the heavy quark is
also at rest.
The first of this type
is due to the QCD interactions with gluons and light quarks
in the heavy meson.  The correction
factor is\cite{neubert,fnl} $\bar{\Lambda} / 2 m_q$, where $\bar{\Lambda}$ is
the mass difference between the heavy meson and heavy quark.  For a
conservatively low value of $m_c = 1.3$ GeV, this correction is roughly
$15 \%$ for the $D$ meson.

The second type of HQET correction
occurs when the momentum transfer between initial and final
mesons is large.  Then, the emission of a hard gluon by
the heavy quark before its decay can give the heavy quark a large velocity
with respect to the meson rest frame.  The naive estimate of this
effect is\cite{neubert,blnn,fnl} $( v\cdot \pp )/ 2 m_q \approx 1/4$
at zero $q^2$.  However, studies of these QCD corrections
indicate that the problem caused by the leading gluon exchange
diagram is strongly suppressed by an order of
magnitude\cite{bd,isgur}.  We have used the perturbative
QCD calculation of sec.~3 to estimate these corrections.  They are
of the form
\begin{equation}
\label{correction}
\Delta \theta_1^{(b)} = - \Delta \theta_2^{(b)} = {w\ \mkay\over m_B} C \ ,
\end{equation}
where
\begin{equation}
C = \int_0^{1-\epsilon} dy {\phi^\dagger (y)\over \Delta_Q D_G} \ .
\end{equation}
This correction comes from from the diagram of fig.~(2a).
We evaluated it for $\phi = \phi_1$ as defined in sec.~3, and
plot the the ratios $\Delta\theta_1 /\theta_1$,
$\Delta\theta_2 /\theta_2$, and $\Delta(A + B)/(A+B)$ in fig.~7.
%
\begin{figure}[t]
\begin{center}
\leavevmode
\epsfxsize=4truein
\epsffile{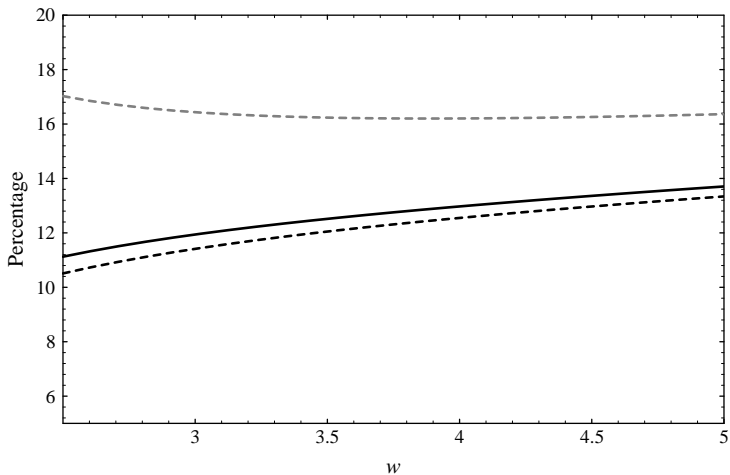}
\end{center}
\caption{
The percentage of heavy quark symmetry violating
$w\  m_{K^*}/m_B$ corrections for $\theta_1$
(dashed gray),  $\theta_2$ (dashed black), and $A + B$ for
the $B \rightarrow K^* \gamma$ process (solid black).
}
\label{figure7}
\end{figure}
As $w$ grows much larger than $\mkay/m_B$, these corrections
actually dominate the asymptotic behavior.  The solid line of
fig.~7 evaluated at $w = w_B$ gives an estimate of these
corrections to \vts\ of about $12\%$.
While the $w_D\ \mkay/m_D$ corrections
are impossible to estimate via perturbation theory because $w_D$ so
close to one,
heavy quark symmetry dictates that they should have the
same form as the $w_B\ \mkay/m_B$
corrections, that is, they are systematically correlated.
The experimental values for the $\theta$'s at $w_D$ includes
these corrections,
while our perturbative
analysis does not.  Therefore these errors systematically
compensate each other in the final result,
so we use $12\%$ uncertainty total, including both corrections.

Another source of HQET violations are perturbative correction which
cannot be included in the definitions of the heavy -- light form factors
$\theta_i$.  From our perturbative calculation, we find that the
dominant source of these corrections,
which are proportional to $(1-\vslash )/2$, are
precisely the same algebraic form as the $w\cdot \mkay/m_B$ corrections
to the $\theta_i$ as discussed above.  It is clearly difficult
to assign a percentage of uncertainty to these corrections,
since we have not calculated the precise way in which they feed into our
calculation of \vts, so we make a naive estimate of
$6\%$ total
uncertainty from them.
This value is smaller than the naive estimate of
$w \mkay/m_B\sim 1/4$, but not by
an order of magnitude as was hoped\cite{bd,isgur}.

We also need to estimate corrections to our perturbative QCD
calculation of the matching conditions given by eqns.~(\ref{ratios}).
That is, we wish to estimate the uncertainty due to the running
of the form factors from $w = w_D$ to $w = w_B$.
It has been suggested in the literature that the $\alpha_s$ contribution
to the form factors may be producing only $10\%$ of the full
result\cite{bd}.  In this case, the matching conditions would
be meaningless.

To check this, we use eqn.~(\ref{Ekappa}) to determine $f_B$.  From the
\dkev\ data and our fits, we find
\begin{equation}
\kappa = f_{K^*}f_B {\sqrt{m_{K^*}m_B} \over m_{K^*}^3}
          {4\pi \alpha_s \over 9} = 0.078 \pm .022(28\%) \ ,
\end{equation}
where we have quoted experimental uncertainties.  To extract
$f_B$ from eqn.~(\ref{Ekappa}), we use $\alpha_s(m_B) = 0.20$, and
$f_{K^*} = 212$ MeV from QCD sum rules\cite{czz}.
This sum rule estimate of $f_{K^*}$ is
supposedly  is good to about $20\%$.  This yields
\begin{equation}
f_B \approx 420\ {\rm MeV}.
\end{equation}
This is a rather large value of $f_B$, when compared to
numerous lattice estimates\cite{pdbook} of $200$ to $300$ MeV.
However, it is not an order of magnitude larger than these values,
which would be the case if the $\alpha_s$ corrections contributed
only $10\%$ to the full result for the form factors.
In addition, the calculation of $f_B$ done here assumes the leading HQET
relations between the $D$ and $B$ meson systems.  Decay constants
are well known to receive large $1/m_Q$ corrections in the
HQET\cite{neubert}, and these should be taken into account before
a real prediction of $f_B$ is made via \dkev\ and \bkg\ data.
A final note on this topic is that there is new
evidence that the lattice calculations may
systematically underestimate heavy quark decay constants.
The recently measured\cite{fds} $f_{D_s}$ decay constant's
central value is about 1.5 times the central values of the
lattice estimates.
We conclude that the order $\alpha_s$ calculation is indeed
dominating the form factor calculation at $w = w_B$.  And a more
proper way of calculating $f_B$ would be to include all
$1/m_B$ corrections in the HQET.

Since the normalization of the form factors is set
by experiment in our case, the error in $f_B$ is not directly
correlated with error in \vts. Our input for \vts\ from
perturbative QCD are the
ratios $r_1, r_2, r_3$ defined in equation (\ref{ratios}).  We currently
have no systematic way of estimating their uncertainties due to
higher order corrections, other that to apply that standard
$\alpha_s / \pi$ rule for the next to leading order correction.
This yields a naively small error estimate of $7\%$.
Clearly, an explicit $\alpha^2_s$ calculation of the type given by
Field et.~al.\cite{field} would pin this uncertainty down further.
The uncertainty in $\alpha_s(m_B)$ is about $10\%$.
Our uncertainty in the soft physics which goes into the
perturbative calculation can also be estimated.  The peaking approximation
for the heavy meson distribution amplitude yields about $\lbar/m_B = 10\%$,
and the uncertainty in the $K^*$ wavefunction about $\mkay/m_B = 17\%$.

In addition, our perturbative QCD calculation is sensitive to the
IR cutoff used to regulate our perturbative momentum fraction
integrals.  This is the standard ``solution'' to the IR problem
in heavy -- light systems as discussed in the
literature\cite{bd,kk,shb}.  This aspect of the calculation
definitively needs to be improved to given more confidence to the
application of perturbative QCD to heavy quark systems.
(Our philosophy in this paper has been to apply this ``well known'' QCD
technology.)
One way of estimating uncertainty
due to this ambiguity is to perform the calculation for
various distribution amplitudes.  While the results were not
displayed, we found that the $\phi_{czz}$ distribution amplitude
gives  essentially the same result for $\vts$,
to within about $10\%$ uncertainty.  The similarity of the
result is already evident from fig.~3.
Finally, theoretical uncertainty from the top quark mass adds
about $8\%$.

We combine the theoretical uncertainty into two parts.
The first part consists of uncertainty due to corrections in HQET and the
top quark mass.  They give about $22\%$ when added in quadrature, with
the dominant contributions comming from $w\cdot\mkay/m_Q$ type corrections.
The second type of theoretical uncertainties come from running
the form factors from $w_D$ to $w_B$ via the perturbative QCD matching
conditions.  We estimate uncertainty of about $23\%$ from perturbative
QCD uncertainties.  Clearly, these are very difficult to estimate
because we don't know how the QCD corrections feed into
the matching conditions eqns.~(\ref{ratios}). However, because we
are using perturbative QCD rather than simply making
a pole or dipole ansatz for the form factors,
we can make an estimate that can be systematically improved as
the perturbative calculations become more sophisticated.
Hence we find a total estimated theoretical uncertainty of
about $32\%$ from all sources when added in quadrature.

\section{summary and conclusions}

We find a value for the mixing angle of about
\begin{equation}
\vts = 0.035 \pm .010 \pm .011 (28\% + 32\%)
\end{equation}
where the second one $\sigma$ uncertainty is a naive estimate of
total theoretical errors. It is naive because it is an estimate
of higher order corrections that we have not calculated.
We believe that this factor can be substantially
reduced by further work on
$1/m_Q$ HQET corrections to the heavy quark matching
between $D$ and $B$ systems, and an order $\alpha_s^2$ calculation
of the perturbative QCD matching conditions for the heavy -- light
form factors at $w = w_B$.  The most disturbing theoretical
uncertainty seems, at this point,
to be the IR sensitivity of the perturbative QCD calculation,
which we have estimated to be $10\%$.

Our value compares very well with the standard model
result\cite{pdbook} of $\vts = 0.041 \pm .007(17\%)$, that is,
our result is consistent with three generations of quarks,
and the standard model contributions to the $b \rightarrow s \gamma$
penguin.

We can compare our result with recent lattice
estimates of the hadronic tensor form factor for \bkg\
decay.
The UKQCD group\cite{ukqcd} estimates $2 T_1 = |A + B| = .58^{+48}_{-56}$,
and with an additional assumption of spectator mass independence,
$|A + B| = .60^{+20}_{-14}$.
Bernard {\it et.~al.}
\cite{bernard}
estimate $|A + B| = .40^{+04}_{-.12}$ using
a pole form ansatz.  These results should be compared
to our result of $|A + B| = 0.78 \pm .16 \pm .25\ (21 \pm 32)\%$,
where we quote separately experimental and theoretical uncertainty.
(There is less
experimental uncertainty than for \vts\ since we
do not require \bkg\ data for this parameter).

The QCD sum rule technique has also been applied to
help determine the \bkg\ form factors\cite{cdnp}, and
a single pole ansatz, with no subtractions, has been
recently used\cite{narduli}.
As discussed in sec.~2, this ansatz is hard to reconcile
with leading order HQET, so that  $\lbar / m_B$ corrections play
a significant role in
determining the form factors by this method. However, both of
these results are in rough agreement with ours.

As part of a self consistency check on the perturbative
part of our analysis, we find the $B$ meson decay constant
$f_B \approx 420 $ MeV with $28\% \pm 32\% \pm 20\%$
uncertainty, where the last $20\%$ is the generic uncertainty in the
$f_{K^*}$ decay constant from QCD sum rules.  While our value
of $\vts$ is {\it not} directly correlated to this value, since
we fix the normalization of our form factors from experiment,
this value does indicate that the perturbative QCD calculation is
correctly estimating the exclusive decay rate at large
meson - meson recoil.

\section*{Acknowledgments}

We would like to thank Rick Field, Pierre Ramond, Charles Thorn and
Ariel Zhitnitsky for useful conversations.
We are particularly grateful to Ben Grinstein for stimulating
our interest in this calculation and for many useful discussions.
P.~G. would also like to
thank the Physics department of
Southern Methodist University for hospitality while this
work was completed.
This work has been supported by SSC Fellowship \#FCFY9318 (P.G.),
the Ministerio de Educaci\'on y Ciencia of Spain (M.Ma.), and
U.S.~Department of Energy grant DE-FG05-86ER-40272 (M.McG., P.G.).

\section*{Appendix A: Calculation of integrals $I_Q$ and $I_s$}

The QCD calculation for $\theta_1$, $\theta_2$ and $\theta_3$
is written in terms of two integrals over the light quark
momentum fraction $y$. These integrals $I_Q$ and $I_s$
arise from the Feynman diagrams containing a heavy or strange
quark propagator, figure 2a and 2b respectively. They are
given by:
$$I_Q= \int_{\epsilon}^1 d\bar{y} \bar{\phi}^{\dagger}_{K^*}(\bar{y})
{m_{K^*}^3 \over \Delta_Q D_g}$$
$$I_s= \int_{\epsilon}^1 d\bar{y} \bar{\phi}^{\dagger}_{K^*}(\bar{y})
{m_{K^*}^3 \over \Delta_s D_g}$$
where $\Delta_Q$, $D_s$ and $D_g$ are given in eqns.~(26) and
$\bar{y}= 1-y$. It
is convenient to rewrite these integrals in terms of
\begin{equation}
I_1 = -2 w I_Q =\int_{\epsilon}^1 d\bar{y}{ \bar{\phi}^{\dagger}_{K^*}
(\bar{y})
\over (\bar{y} - {\bar{\Lambda} \over m_{K^*}w})
(\bar{y}^2 + {\bar{\Lambda}^2 \over m_{K^*}^2}
- 2{\Lambda \over m_{K^*}}w\bar{y} )
}
\end{equation}
and
\begin{equation}
I_2 = {\Delta_s \over m_{K^*}}I_s =\int_{\epsilon}^1 d\bar{y}
{ \bar{\phi}^{\dagger}_{K^*}(\bar{y})
\over \bar{y}^2 + {\bar{\Lambda}^2 \over m_{K^*}^2}
- 2{\Lambda \over m_{K^*}}w\bar{y} }
\end{equation}
By partial fractions these integrals can be written in terms of a
single function
\begin{mathletters}
\begin{eqnarray}
I_1 & = & {1 \over a_1-a_3}{1\over a_1- a_2}(f(a_1) - f(a_3))
- {1\over a_1 - a_2}I_2 \ , \\
I_2 & = & {1\over a_2 - a_3}(f(a_2)-f(a_3)) \ ,
\end{eqnarray}
\end{mathletters}
where
\begin{mathletters}
\begin{eqnarray}
a_1 & = & {\bar{\Lambda} \over m_{K^*}w}\ , \\
a_2 & = &{\bar{\Lambda} \over m_{K^*}}(w + \sqrt{w^2-1})\ , \\
a_3 & = & {\bar{\Lambda} \over m_{K^*}}(w - \sqrt{w^2-1}) \ ,
\end{eqnarray}
\end{mathletters}
gives the position of possible poles in $1-y$, (before the Drell-Yan-West
region has been cutoff), and the function
$f(a_i)$ is defined by
\begin{equation}
f(a_i) =  \int_{\epsilon}^1 d\bar{y}{ \bar{\phi}^{\dagger}_{K^*}(\bar{y})
\over \bar{y}- a_i} \ .
\end{equation}

The value of $f(a_i)$ involves the cutoff prescription $\epsilon$ and
the form of the $K{^*}$ wave function.
In this paper we have employed
the $w$ dependent cutoff
$\epsilon = {\bar{\Lambda} \over m_{K^*}\sqrt{w^2-1}}$. If the $K^{*}$
wave function is taken to be $\bar{\phi}^{\dagger}_{K^*}(\bar{y}) =
6 y (1-y)$, then
\begin{equation}
f(a_i) = 3 - 6a_i + 3\epsilon^2 + 6(a_i-1)\epsilon
+6a_i(1-a_i)\ln{|1-a_i|} -6a_i(1-a_i)\ln{|\epsilon-a_i|}
\end{equation}
Note that this function, together with $I_Q$, $I_s$ and $\theta_1$,
$\theta_2$, $\theta_3$, depends only on ${\bar{\Lambda}\over m_{K^*}}$
and $w$, and is independent of $m_Q$ as required by heavy quark symmetry.
The explicit form for the $\theta_i$ involves logarithms the largest
of which goes like ${\ln{w} \over w^2}$ as found in the asymptotic
expansions given by eqn.~(30).

\end{document}